%% file: icl_sim.tex
\documentclass[apj]{emulateapj}
\usepackage{apjfonts}

\newcommand{\ie}{i.e.,\ } \newcommand{\eg}{e.g.,\ }
 \newcommand{\etal}{et~al.\ }
\newcommand{\ltsima}{$\; \buildrel < \over \sim \;$}
\newcommand{\simlt}{\lower.5ex\hbox{\ltsima}} \newcommand{\gtsima}{$\;
\buildrel > \over \sim \;$}
\newcommand{\simgt}{\lower.5ex\hbox{\gtsima}}  \newcommand{\magsec}{mag/arcsec$^2$}
\newcommand{\ficl}{$f_{ICL}$} \newcommand{\fficl}{$f_{FICL}$}
\newcommand{\dficl}{$\Delta f_{ICL}$} \newcommand{\dfficl}{$\Delta
f_{FICL}$} \newcommand{\lcdm}{$\Lambda\mbox{CDM}$\ }
\newcommand{\muv}{$\mu_V$}

\begin{document}

\title{The Formation and Evolution of Intracluster Light}

\author{Craig S. Rudick, J. Christopher Mihos, and Cameron
McBride\altaffilmark{1}} \email{craig@fafnir.astr.cwru.edu,
mihos@case.edu, ckm8@pitt.edu} \affil{Department of Astronomy, Case
Western Reserve University, 10900 Euclid Ave, Cleveland, OH 44106}
\altaffiltext{1}{Now in the Department of Physics and Astronomy,
University of Pittsburgh.}

\begin{abstract}

Using $N$-body simulations, we have modeled the production and
evolution of diffuse, low surface brightness intracluster light (ICL)
in three simulated galaxy clusters. Using an observational definition
of ICL to be luminosity at a surface brightness \muv$>26.5$ \magsec,
we have found that the fraction of cluster luminosity contained in ICL
generally increases as clusters evolve, although there are large
deviations from this trend over short timescales, including sustained
periods of decreasing ICL luminosity.  Most ICL luminosity increases
come in short, discrete events which are highly correlated with group
accretion events within the cluster.  In evolved clusters we find that
$\approx 10-15$\% of the clusters' luminosity is at ICL surface
brightness.  The morphological structure of the ICL changes with time,
evolving from a complex of filaments and small-scale, relatively high
surface brightness features early in a cluster's history, to a more
diffuse and amorphous cluster-scale ICL envelope at later times.
Finally, we also see a correlation between the evolution of ICL at
different surface brightnesses, including a time delay between the
evolution of faint and extremely faint surface brightness features
which is traced to the differing dynamical timescales in the group and
cluster environment.

\end{abstract}


\section{Introduction}

Diffuse, intracluster starlight (ICL) consists of stars in galaxy
clusters which have been gravitationally stripped from their parent
galaxies via cluster galaxies interacting with other galaxies or with
the cluster potential.  As a product of the dynamical interactions
within the cluster, the ICL has the potential to reveal a great deal
of information about the cluster's accretion history and evolutionary
state, as well as the mass distribution of cluster galaxies and the
cluster as a whole.  The quantity, morphology, and kinematics of the
ICL each hold potentially useful information on the cluster's
evolution, and processes affecting individual galaxies can be traced
using individual ICL streams.

Observationally, ICL has been detected in numerous galaxy clusters in
the local universe through broadband imaging. Early measurements of
ICL (Zwicky 1951; Oemler 1973; Gudehus 1989) were extremely difficult
due to the very low surface brightness of ICL features, at less than
1\% of sky brightness.  The advent of modern techniques for precision
CCD photometry has caused a resurgence of interest in the subject
(e.g., Uson et al. 1991; Gonzalez et al. 2000; Feldmeier et al. 2002,
2004; Da Rocha \& de Oliveira 2005).  Dozens of clusters from the very
nearby universe, such as Virgo (Mihos et al. 2005) and Coma (Adami et
al. 2005), to the intermediate redshift universe
(V{\'i}lchez-G{\'o}mez et al. 1994; Zibetti et al. 2005) have now been
studied for signs of intracluster luminosity.  Most studies estimate
that the ICL comprises $\approx 10-40$\% of the clusters' luminosity,
and it is now thought that ICL is a ubiquitous feature of evolved
galaxy clusters.

The ICL can also be studied using discrete tracers of stellar
populations, such as intracluster planetary nebulae (IPNe)
(e.g. Arnaboldi et al. 1996, Feldmeier et al. 1998, Aguerri et
al. 2005), red giants (Durrell et al. 2002), novae (Neill et al 2005),
and supernovae (Gal-Yam et al. 2003).  These studies of individual
intracluster stars are generally limited to the nearby universe due to
the extreme difficulty in detecting these objects.  However, estimates
of ICL luminosity using these methods are in broad agreement with
those found through broadband imaging.  Of these stellar tracers, IPNe
are especially interesting as radial velocity measurements of IPNe
have opened up the possibility of studying the kinematics of the ICL
(Arnaboldi et al. 2004; Gerhard et al. 2005).

While the existence of ICL is now well established, questions remain
surrounding the details of when, where, and how the ICL is formed.  A
variety of scenarios have been put forward to explain the production
of ICL: stripping during the initial collapse of the cluster (\eg
Merrit 1984); stripping of galaxies by an established cluster
potential (Byrd \& Valtonen 1990; Gnedin 2003); stripping within
galaxy groups accreting onto the cluster (Mihos 2004); and stripping
from high speed encounters between cluster galaxies (Moore et
al. 1996). Indeed in the complex environment of a collapsing and
accreting galaxy cluster, all these processes likely contribute to the
overall production of ICL, indicating that full $N$-body simulations
of the cluster evolution are needed to probe ICL formation.

Motivated by these considerations, several recent studies have focused
on simulating ICL production in a \lcdm universe, using either dark
matter simulations with tracer particles (\eg Napolitano \etal 2003)
or full hydrodynamical galaxy formation models (\eg Murante \etal
2004; Willman \etal 2004; Sommer-Larsen \etal 2005) and have each
found that at $z=0$, at least 10\% of the clusters' stars were unbound
to any one galaxy, in line with current observations.  Napolitano
\etal (2003), Sommer-Larsen \etal (2005), and Willman \etal (2004)
each study the kinematic distribution of the unbound stars and find
significant kinematic substructure.  Willman \etal (2004) make
detailed simulated observations of IPNe and show that they can be used
to trace specific tidal features.

Each of these studies, however, focuses on ICL defined as stars which
are unbound to any cluster galaxy.  In general, the specific binding
energy of individual stars is not a readily observable feature of the
ICL.  While IPNe are already proving useful in determining the
kinematic structure of the ICL in nearby clusters, most observational
studies continue to rely on broadband imaging to quantify the
properties of the ICL. To make the connection between observations and
simulations more direct, we adopt here a more observational definition
of ICL: luminosity below a given surface brightness threshold. Using
this definition, we study the ICL properties of three simulated galaxy
clusters in a \lcdm universe.  We describe the simulation and
artificial imaging techniques in Section 2, and Section 3 explores the
mechanisms driving the evolution of the ICL.

\section{Simulations}

To simulate the formation of intracluster light we use a technique
similar to that described by Dubinski (1998), but updated to reflect a
modern \lcdm cosmology, a mix of both spiral and elliptical galaxy
types, and a halo subdivision scheme to avoid overly massive galaxies
in the simulation. Full details of the simulation techniques will be
given elsewhere (Mihos \etal in preparation); we highlight the most
important features here.

To create the initial conditions for the cluster simulations, we first
run a $N=256^3$ 50x50x50 Mpc $\Lambda=0.7$, $\Omega_M=0.3$, $H_0=70$
cosmological dark matter simulation from $z=50$ to $z=0$, at which
point collapsed clusters with masses $O(10^{14}) M_\sun$ are chosen to
re-simulate at higher resolution. For each cluster, we identify
individual halos at $z=2$ which are destined to end up within the
$z=0$ cluster, and insert higher resolution collisionless galaxy
models into these halos.  Given a halo mass, we first decide the
number of galaxies to insert using a Monte Carlo technique based on
the ``halo occupancy distribution'' (HOD) formalism (Berlind \&
Weinberg 2002), in which the number of galaxies inserted into a halo
of mass $M$ is given by
 
\[ N_{gal} = \left\{ \begin{array}{ccc}
                      \mbox{int($(M/M_{MW})^\alpha$)} & \mbox{if} &
                      \mbox{$M/M_{MW}>1$} \\ 1 & \mbox{if} &
                      \mbox{$0.1 < M/M_{MW} < 1$} \\ 0 & \mbox{if} &
                      \mbox{$M/M_{MW}<0.1$}
                     \end{array}
             \right. \]

\noindent where the Milky Way mass is taken to be $M_{MW}=6\times
10^{11}\ M_\sun$ and the HOD index is given by $\alpha=0.7$.

For each substituted halo, the original dark matter particles
(referred to as ``cosmological dark matter'') are sorted by
environmental density and the $70\%$ most dense particles are excised
from the simulation, to be replaced with high resolution galaxy
models. The remaining cosmological dark matter particles are left
intact and fill the volume around each substituted galaxy model as an
extended dark matter halo (for a single galaxy within a halo) or
common dark matter envelope (for multiple galaxies within a halo).
When multiple galaxies are substituted, the individual masses are
drawn from a power law mass function, subject to the constraint that
the total mass of all galaxies must be equal to the mass being
substituted. The final galaxy mass function is set by the convolution
of the underlying halo mass function, the HOD function, and the
power-law function for substituting multiple galaxies, and shows a
Schecter-like high mass cutoff (from the rarity of high mass halos)
and behaves like a power law (slope $\sim$ -1) at lower masses.  For
halos with multiple galaxies inserted, the galaxy positions and
velocities within the halo are determined by statistically sampling
the densest subpeaks of the original dark matter halo.

When substituting galaxies, two types of models are used: a disk model
in which the stars follow a composite exponential disk plus Hernquist
(1990) bulge (with bulge-to-disk ratio of 1:5), and an elliptical
galaxy model where the stars have a pure Hernquist (1990)
distribution. The disk galaxy model is built using the prescription of
Hernquist (1993), while the elliptical galaxy is built in a similar
fashion, by omitting the disk component and using a massive spheroid
instead. Both models are embedded in isothermal dark halos (referred
to as ``galaxy dark matter''); these halos possess a constant density
central core inside one disk scale length (for the disk galaxy model)
or Hernquist scale radii (for the elliptical galaxy model), and a
maximum extent of 10 scale lengths or scale radii.  In these models,
the galaxy dark matter mass is six times the stellar mass; when
coupled with the extended cosmological dark matter mass left in the
original halo, this gives the galaxies a dark-to-stellar mass ratio of
10:1. The galaxy models have been built in dynamical equilibrium (see
Hernquist 1993 for details), and have been evolved in isolation for
thirty half-mass rotation periods to ensure stability. We also track
the detailed evolution of individual, relatively isolated galaxies
within the full cluster simulation and see no evidence of any
instability which might be caused by the insertion process itself.

The choice of whether to insert a spiral or elliptical model into any
given halo is based on the local galaxy density, following the
morphology-density relationship of Dressler (1980). We first calculate
the local galaxy number density around each galaxy, then assign a
galaxy type based on this galaxy density, where the fraction of
ellipticals rises from 10\% in the lowest density quartile to 40\% in
the upper quartile. For disk galaxy models, the orientation of the
disk plane is randomly chosen. Given a galaxy mass, galaxy sizes are
then scaled by $M^{0.5}$ and velocities are scaled by $M^{0.25}$. This
scaling preserves surface density and keeps spiral galaxies on a
Tully-Fisher-like relationship. Applied to elliptical galaxies, this
scaling introduces a slight tilt to the Fundamental Plane, but this
tilt is modest given the limited range of initial galaxy masses used.

We illustrate the initialization process with statistics from one of
our clusters.  In this cluster, a total of 121 high resolution galaxy
models (91 spirals and 30 ellipticals) were inserted into 80 dark
halos.  The number of particles used in each galaxy model scales with
total galaxy mass, so that all star particles have the same mass. In
total, the simulation consists of approximately 10 million particles:
5.4 million star particles, 2.4 million galaxy dark halo particles,
and 2.2 million cosmological dark halo particles. The
Plummer-equivalent gravitational softening length of the particles is
set at 300 pc for stars, 1.4 kpc for galaxy dark matter, 10 kpc for
cosmological dark matter, and 50 kpc for cosmological dark matter well
outside the cluster (at $r>$ 8 Mpc). Initialized at a redshift of
$z=2$, the cluster is then evolved to $z=0$ using the $N$-body code
GADGET (Springel et al. 2001).

In order to test the effects of the chosen galaxy substitution
redshift, we have re-initialized one of our cluster at $z=3$ and
compared its evolution to that of the standard cluster with
substitution at $z=2$.  Due to the hierarchical nature of the growth
of structure in the universe, the cluster at higher redshift contains
dark matter halos which tend to be smaller than those found at lower
redshift.  Thus, in the high redshift cluster, fewer galaxies are
inserted because fewer dark matter halos meet our minimum substitution
mass, and the galaxies which are inserted tend to be systematically
smaller due to the lower dark halo masses.  With these caveats in
mind, we find that our results are quite similar in both simulations.
All further results presented in this paper refer to clusters in which
galaxies were substituted at $z=2$.

We have simulated three such clusters, referred to throughout this
paper as clusters C1, C2, and C3.  Each cluster is a separate, unique
object taken from the same cosmological simulation, selected to be of
approximately the same mass.  Thus, there are no systematic
differences between the clusters, only variations in the initial mass
distribution resulting from the cosmic variance of mass distribution
within the simulation.  Parameters of the individual clusters can be
found in Table \ref{tab:ctable}, including $R_{200}$ (the radius
within which the density of the cluster is 200 times the critical
density), $M_{200}$ (the mass enclosed in $R_{200}$), the number of
galaxies initially inserted at $z=2$, and the number of galaxies
remaining at $z=0$.

To be sure, our simulation scheme is a simple caricature of the very
complex process of galaxy cluster evolution. The use of collisionless
models neglects the effects of gas accretion, ram pressure stripping,
and star formation, all of which could act to alter the relative
spatial distribution of stars in the evolving cluster.  However, by
concentrating on gravitational dynamics alone, we avoid the
significant uncertainties involved with modeling gas physics, star
formation, and feedback, and isolate the role that gravitational
stripping plays in the creation of intracluster light.  Another issue
is our galaxy substitution scheme. Our low mass substitution limit
corresponds to 10\% of the Milky Way mass, meaning that our
simulations are essentially focusing on the evolution of fairly
massive galaxies; the stripping and destruction of low mass dwarf
galaxies are not captured in our simulations. This may result in a
systematic underestimate of the ICL fraction in our simulations, since
the effects of tidal stripping are enhanced for dwarf galaxies (\eg
Moore \etal 1998). For a Schecter luminosity function with faint end
slope $\alpha=-1$, galaxies below 0.1 Milky Way masses contribute
about 10\% of the total luminosity, placing an upper limit on the
amount of ICL that we could be missing due to our substitution limit.
Also, by choosing a redshift of $z=2$ for substitution, we miss halos
which may be just below the substitution mass limit at $z=2$ but later
grow to exceed the substitution limit. Because we do not insert
galaxies in these halos, we miss their contribution to the ICL;
however, since their host halos remain intact, they still contribute
to the stripping of other galaxies, so that the dynamics of the
substituted galaxies remains accurately tracked. With these caveats in
mind, we now proceed to study the properties of diffuse light in these
cluster simulations.

\subsection{Simulated Images}

To create the simulated cluster images, we project onto two dimensions
the particle distribution, creating an image of $4k\times4k$
pixels,with each pixel 800 pc on a side.  For each cluster we imaged
80 snapshots equally spaced in $\log$(expansion factor) from $z=2$ to
$z=0$.  We have labeled all simulated images in units of time, such
that $t=1.0$ is the current age of the universe.

All simulated images and analyses were carried out using a fixed
viewing angle; the effect of varying viewing angle on the derived
properties of the ICL is examined below.  In order to create a
realistic, continuous mass distribution from our discrete particle
data, we Gaussian smooth the particles in an SPH-like way, where each
particle has smoothing length $\sigma$ which is proportional to the
local three-dimensional particle density. The proportionality constant
determining how the width of the Gaussian is scaled with density was
determined by visual inspection of the images, in order to balance the
competing effects of smoothing the particles enough to remove their
discreteness, while not over-smoothing which can destroy legitimate
spatial structures. The maximum distance to which any particle was
smoothed was limited to the shorter of $4\sigma$ or 400 kpc, and the
final results are not highly sensitive to the detailed choice of the
smoothing parameter value, over a wide range of qualitatively
reasonable values.

Converting the projected mass distribution into a luminosity
distribution requires applying a mass-to-light ratio ($M/L$) based on
the evolutionary state of the stellar population.  This, however,
requires assumptions about the star formation rate, the IMF, stellar
evolutionary tracks, etc., none of which we have modeled in our
$N$-body simulations.  Additionally, an evolving $M/L$ would have the
effect of obscuring many aspects of the dynamical evolution of the
cluster, by conflating dynamical changes with stellar evolutionary
effects. For simplicity, we have instead chosen to use a fixed stellar
mass to light ratio of $5M_{\odot}/L_{\odot}$ for all of our images, a
characteristic value for the $V$-band luminosity of an evolved stellar
population such as we expect to comprise the ICL of a galaxy cluster
in the local universe.  One of the ramifications of using this
constant $M/L$ is that we are not attempting to simulate observations
of the clusters at distant redshifts, but we are simulating
observations of the clusters at a given point in their dynamical
evolution, as they would appear at $z=0$.  This makes our simulated
observations more relevant to observations of the local universe where
there are many galaxy clusters at a variety of dynamical ages.

As noted in \S 1, many different working definitions have been
proposed for the ICL, including material which is unbound to any
cluster galaxy, material at very low surface brightness, and diffuse
material which is morphologically distinct from extended envelopes
around galaxies. Here we use a definition which is both quantifiable
and observationally tractable: luminosity which has a $V$-band surface
brightness fainter than 26.5 \magsec.  This limit coincides with the
Holmberg radius (Holmberg 1958), a commonly used surface brightness
limit defining the isophotal size of galaxies.  This choice is further
motivated by inspection of both our simulated images and of deep
imaging of galaxy clusters (Mihos \etal 2005; Feldmeier \etal 2002),
where \muv$= 26.5$ is approximately the surface brightness limit where
the isophotal contours no longer simply outline those at higher
galaxian surface brightness.  That is, this limit appears to be the
surface brightness where the diffuse light takes on its own
morphology, as opposed to being a continuation of the galaxies'
extended profiles. In section \S \ref{sec:ficl} we also examine the
evolution of ICL under a fainter choice of ICL threshold: \muv$=30$.

Figure \ref{fig:example} shows an example of one of our simulated
images, color coded by $V$-band surface brightness.  Black represents
all luminosity at \muv$<26.0$ \magsec.  Each other color represents a
bin of one \magsec: red is \muv$=26.0-27.0$, orange is
\muv$=27.0-28.0$, etc.  Our images do not show luminosity which is
fainter than \muv$=33.0$ \magsec.  This same color scheme will be used
in all simulated images shown in this paper.  Also, at the top right
of Figure \ref{fig:example}, the distance scale of the image in
physical units is shown.

Because our working definition of ICL is based on surface brightness,
a projected property of the cluster, it is possible that our results
could be sensitive to the choice of viewing angle. To test this
effect, we measured the distribution of luminosity as a function of
surface brightness for 7 different viewing angles of each cluster at 3
different times. The 7 viewing angles used were our standard viewing
angle used for analysis, plus rotations of $\pi/4$, $\pi/2$, and
$3\pi/4$ in both the $\theta$ and $\phi$ directions. Figure
\ref{fig:rotlum} shows the fraction of the luminosity in each cluster
which is in surface brightness bins of 1 \magsec\ at $z=1.0$
($t=0.45$), $z=0.5$ ($t=0.64$), and $z=0.0$ ($t=1.0$).  The points are
the mean values and the error bars indicate the minimum and maximum
values found in those 7 viewing angles.  Over the timescales shown,
the min-max error bars tend to be smaller than the separation between
the means.  This indicates that the physics of cluster evolution
dominates over projection effects in the observed evolution of the
clusters.  The only case where the cluster evolution does not dominate
projection effects is the evolution of C1 from $z=0.5$ to $z=0$.  We
will see in \S \ref{sec:clusters}, however, that this is the result of
the specific evolution of this cluster.

Figure \ref{fig:rotlum} also shows that as the cluster evolves, the
fraction of luminosity at high surface brightness decreases while the
fraction of luminosity at low surface brightness increases.  This
indicates that luminosity is being transferred from high to low
surface brightness.  The fraction of cluster luminosity at ICL surface
brightness clearly increases with time.  The mechanism driving this
redistribution of luminosity is the stripping of material from
galaxies to form the ICL.  We will explore the production of ICL
luminosity in more detail in \S \ref{sec:prodicl}

\section{Evolution and Production of ICL} \label{sec:prodicl}

A full understanding of the nature of ICL requires measuring its
evolution and exploring the mechanisms which drive this evolution.  To
this end, we have measured the luminosity of the ICL throughout the
evolution of each of our clusters.  Furthermore, we have searched for
correlations between the changes in ICL luminosity and specific events
in the clusters' evolution.

During the formation and evolution of galaxy clusters, a variety of
gravitational processes act to liberate stars from galaxies, leading
to the formation of the intracluster light. Slow interactions between
galaxies in infalling groups (Mihos 2004), repeated fast interactions
between galaxies inside the cluster (Moore et al 1996), tidal
stripping of galaxies by the cluster potential (Byrd \& Valtonen
1990), and interactions between galaxies and more massive
substructures within the cluster (Gnedin 2003) all operate in concert
to remove stars from their cluster galaxies.  In
cosmologically-generated simulations such as those presented here,
where all these process are at work, it is difficult to uniquely
disentangle any one of these effects from the overall evolution of the
system, and we make no attempt to do so here. Instead, since all these
processes act to strip material from galaxies, we will generically
refer to these combined processes as "gravitational stripping" and
focus more on their net (observational) result rather than on the
differing details of the gravitational dynamics involved in each one.

\subsection{Evolution of ICL Luminosity} \label{sec:evicl}

The top portion of Figure \ref{fig:delta} shows the fraction of the
total luminosity in the cluster which is at \muv$>26.5$ (\ficl) as a
function of time for all three clusters.  The obvious trend is that
the luminosity of the ICL increases with time in all the clusters, as
material is stripped from galaxies as they first interact in the group
environment, then later aggregate to form the cluster. On shorter
timescales, however, there are significant deviations from the overall
trend of increasing ICL luminosity fraction.  The middle section of
Figure \ref{fig:delta} shows the fractional change in ICL luminosity
per unit time (\dficl\ function), calculated by:
\begin{equation}
\Delta f_{ICL,i} = \left( \frac{f_{ICL,i} - f_{ICL,i-1}}{f_{ICL,i-1}}
\right) \left( t_i - t_{i-1} \right) ^{-1}
\end{equation}
There is a significant dispersion on short timescales in the \dficl\
function.  Each cluster experiences several significant events of
\emph{decreasing} ICL luminosity, with the major increases coming in
short, large events.  In \S\ref{sec:clusters} we will show that these
large increases correlate very strongly with major accretion or
collisional events between galaxies and galaxy groups within the
cluster, while the decreases are caused by material initially
collapsing into higher density environments, thus temporarily raising
its surface brightness.

The bottom plot in Figure \ref{fig:delta} shows the running mean
($\hat{\mu}$) and standard deviation ($\hat{\sigma}$) of the \dficl\
function, using a running bin of ten time points.  The time plotted is
the average of the time points in each bin.  In each cluster, there is
at least one point after $t=0.6$ where $\hat{\mu}$ is negative.  For
even one point of the running mean to be negative, the clusters must
be losing ICL luminosity on average over a significant period of their
evolution (the size of one running bin).  The width of the running bin
in time units changes due to the logarithmic spacing of the data
points, however after $t=0.6$ the minimum size of one bin is 0.075, or
at least 11\% of the cluster's evolution.  This variability in the
cluster's ICL luminosity evidences the complexity of the processes
which produce the intracluster light.

\subsubsection{Initial Evolution of Cluster ICL} \label{sec:initial}
One clear feature in Figure \ref{fig:delta} is the extremely similar
behavior demonstrated by all three clusters at very early times ($t <
0.35$).  Each cluster displays a sharp decrease in ICL luminosity
almost immediately at the start of the simulation, followed by a
significant increase over a very short timescale. Because the galaxy
models have been constructed in equilibrium, this behavior is not due
to instabilities with the galaxies themselves.

We have verified this fact by visually examining isolated galaxies at
the beginning of the simulation, and find very little evolution in
them until they begin interacting with nearby neighbors. Instead,
visual inspection shows that much of the early evolution in the
cluster ICL is due to prompt interactions between small groups of
galaxies. These groups are not necessarily in virial equilibrium, and
several of them quickly collapse once the simulation is started by
first compressing the galaxies, then tidally striping material out to
low surface brightness.  Additionally, we have constructed the
\dficl\ function for these isolated systems and find that the structure
of the \dficl\ function for the entire cluster closely matches that of
the the collapsing group environment.  Thus, we find that the early
evolution of the \dficl\ function of the clusters is the result of the
physics resulting from the initial collapse of galaxies in the group
environment.

However, one should remain cautious about over-interpreting the
clusters' behavior in the initial stages of our simulations.  The
initial luminosity distribution of the clusters is not entirely
realistic, due to the somewhat artificial way in which we inserted
isolated, unevolved galaxies into the cluster potential.  Thus, we
shall focus our analysis on the evolution of the clusters after $t
\approx 0.4$, when the physics of cluster evolution should dominate
over effects due to the initialization of the cluster and the
prescription for inserting galaxies into the model.

\subsection{Individual Clusters}\label{sec:clusters}

While all of our model clusters demonstrate similar global trends,
each has a unique dynamical history which manifests itself in the
cluster's luminosity evolution.  By following the evolution of each
cluster individually, we can identify the events which significantly
impact the production of ICL.

\subsubsection{Cluster C1}

The ICL evolution of C1 is characterized by large gains in ICL
luminosity early in its history, between $t=0.4-0.6$, followed by an
extended period of very little activity.  The early gains experienced
by C1 are very similar to those of the other clusters.  The \dficl\
function is quite erratic in this region, showing both large gains and
large losses occurring over short periods of time.  This is further
demonstrated by the $\hat{\sigma}$ of the \dficl\ function in this
range, which is quite large.

Figure \ref{fig:c1} shows a series of simulated images illustrating
the evolution of C1, with a color scale identical to that used in
Figure \ref{fig:example}.  In the early stages of cluster collapse,
the many different galaxies are interacting in small groups, which
later merge together to form the large cluster. With the complexity of
these small-scale interactions taking place throughout the volume, it
is extremely difficult to determine the impact individual events have
on the evolution of ICL.  Similarly complex behavior in the early
stages of the simulations is seen in all of the clusters, as galaxies
first collapse into groups; this generic result illustrates the
importance of interaction in the group environment that begin the
formation of the diffuse ICL well before the cluster as a whole has
collapsed and virialized.

After $t=0.6$, C1's \ficl\ is quite flat, and the $\hat{\sigma}$ of
the \dficl\ function is much smaller than it is before this time.
Only a few percent at most of the cluster luminosity is converted to
ICL from $t=0.6-1$.  This extremely small increase in ICL luminosity
is the result of the specific evolutionary history of the cluster,
seen in Figure \ref{fig:c1}.  By $t=0.6$, three massive galaxy
complexes have clearly developed. While it is difficult to determine
from Figure \ref{fig:c1} the internal structure of these complexes, in
general they are groups dominated by a massive spheroidal galaxy, with
several smaller galaxies in close proximity.  The most striking aspect
of the evolution of C1 at these later times is that these three
massive complexes which formed by $t=0.6$ still exist intact and
relatively unaltered at $t=1$.  In essence, there are no major
accretion or interaction events taking place between the groups, and
without these violent events to quickly add ICL, the ICL fraction
rises only slowly at late times in this cluster.

\subsubsection{Cluster C2}

The early luminosity evolution of C2 is quite similar to that of C1,
as galaxies interact on group scales throughout the volume.  However,
whereas C1's \ficl\ flattens after $t=0.6$, that of C2 continues to
increase substantially with time.  The \dficl\ function of C2 is quite
active throughout its evolution, with its $\hat{\sigma}$ relatively
large at all times.

Figure \ref{fig:c2} shows simulated images of the evolution of C2.
Whereas C1 quickly develops three distinct massive complexes, C2's
mass buildup consists many small groups continually coalescing to form
one very large central mass complex.  At $t=0.6$, when the three mass
complexes of C1 were in place, the major mass concentration in the
upper right of C2 consists of many separate galaxies which are still
in the process of merging.  Additionally, there are three distinct
galaxy groups to the lower left of the main mass which have yet to be
incorporated into the cluster core.  As time progresses, the main mass
concentration continues to consolidate through mergers, while the
other outlying galaxies are each drawn into the concentration in turn.
The large number of accretion events where massive galaxies and groups
fall into the cluster center create the frequent, large increases in
\ficl\ which occur throughout C2's evolution.

\subsubsection{Cluster C3} \label{sec:c3}

The evolution of C3 (Figure \ref{fig:c3}) provides the most dramatic
example that the production of ICL luminosity is driven by merger
events between massive groups of galaxies. As was the case in C1 and
C2, the \ficl\ of C3 is sporadic, but increasing overall, before
$t=0.6$. However, its subsequent evolution is quite unique. At $t=0.6$
there are 4 distinct massive groups which collapse into a single, very
concentrated massive complex at $t=0.7$. The rapidly changing tidal
field of the collapsing cluster, coupled with many small-scale tidal
interactions, leads to significant stripping of the cluster galaxies
and an enormous rise in the \ficl\ of C3 between $t=0.68$ and $t=0.8$,
where the ICL luminosity more than doubles.

It is also interesting to note the significant {\it decrease} in
\ficl\ as this strong ``crash'' of galaxies begins, starting at about
$t=0.6$. The decrease is due to the fact that as the galaxy groups
collapse, the luminosity becomes highly concentrated.  This
concentration of luminosity elevates the surface brightness of much of
the luminosity in the cluster; material below our \muv$>26.5$
definition for ICL is, for a short-time, boosted to brighter, non-ICL
levels.  While projection effects play some role in this phenomenon,
it is also attributable to a real (but transient) rise in the physical
density of luminous mass.  A similar process is responsible for many
of the major ICL luminosity decreases which are observed to precede
major accretion events in the evolution of each of the clusters.

\subsubsection{Increased Time Resolution of an Accretion Event} \label{sec:hires}

To further illustrate that the evolution of the \ficl\ is the result
of interactions between massive galaxies and groups, Figure
\ref{fig:hires} shows one such event from C2 at higher time
resolution.  Figure \ref{fig:hires_plot} shows the corresponding
regions of the \ficl\ and \dficl\ plots.

The major event that Figure \ref{fig:hires} depicts is the accretion
of a small galaxy group (indicated by the arrow) which is below and to
the left of the main body of the cluster at $t=0.86$.  The series of
images clearly shows this group moving through, interacting with, and
exiting from the cluster core, resulting in a large net increase in
ICL luminosity.  Again, however, from $t=0.88$ to $t=0.9$, the cluster
actually loses ICL luminosity as the group moves closer to the cluster
center.  The situation is similar to that described in Section
\ref{sec:c3}, whereby the increased surface brightness of the diffuse
starlight is caused by both a temporary increase in the local density,
as well as the projection of larger amounts of material onto the line
of sight.  As the group begins to exit the cluster center, at $t=0.92$
and 0.93, the ICL luminosity increases slightly, driven largely by the
fact that diffuse light in the cluster and group is no longer boosted
in luminosity simply by projection.  Immediately thereafter, however,
we see larger gains in ICL luminosity at $t=0.94$ and 0.96. The tidal
field of the cluster has stripped material from the galaxy group,
substantially increasing the luminosity at low surface brightness.
This can be most easily seen as the low surface brightness plume
emerging from the small group and extending to the upper right at
$t=0.96$ (and even more extended at $t=1$ in the last panel of Figure
\ref{fig:c2}).

This type of group accretion/interaction event is responsible for
nearly all of the major ICL luminosity increases observed in our
clusters.  Typical events are characterized by group scale
interactions stripping material from individual galaxies and beginning
the production of diffuse light, followed by an accretion of the group
into the denser cluster core which then generates significantly more
ICL luminosity. Colloquially, the strong, close interactions in the
group environment ``soften up'' the galaxies so that the large scale
tidal field of the evolving cluster can more easily strip material
from galaxies and form the ICL (see also Mihos 2004).  The increases
in the \ficl\ function show that ICL luminosity in a cluster which
formed by the accretion of groups is greater than the sum of the ICL
luminosities of the individual groups alone.

\subsection{Evolution of Very Faint ICL (\muv$>30.0$)} \label{sec:ficl}

One of the systematic effects imposed by our definition of ICL is that
the ICL luminosity is dominated by the bright end of the defined
magnitude range (see Figure \ref{fig:rotlum}).  Thus, the above
results are not truly sampling the entire range of ICL surface
brightness.  In order to trace the evolution of more faint ICL
features, we have created another surface brightness cutoff at
\muv$=30.0$ \magsec, and we define luminosity with equal or greater
magnitude to be Faint ICL (FICL).  Note that FICL luminosity is a
subset of ICL luminosity, albeit only at a small fractional level.

The top of Figure \ref{fig:faint} shows both the \dficl\ and \dfficl\
functions, both calculated as described in Equation 1.  The \dfficl\
function shows a very similar behavior to the \dficl\ function, with
changes in the \fficl\ coming in short stochastic events.  However,
while the two functions share many common characteristics, the
evolution of the ICL and FICL are certainly not coincidental.  Using
the procedure described below, we find that certain segments of the
\dfficl\ function are highly correlated with segments of the \dficl\
function, provided that the \dfficl\ segments are shifted backward in
time; \ie the evolution of the FICL lags behind the evolution of the
brighter ICL.  This behavior can be seen in the second row of Figure
\ref{fig:faint}, which overplots the shifted segments of the \dfficl\
function with the \dfficl.

In order to calculate the correlation between the two functions, we
first map the functions onto a regular time grid by fitting them using
a third order polynomial interpolation (Press \etal 1986), evaluated
at regular intervals of $dt=0.001$.  Note that the lines in Figure
\ref{fig:faint} follow the interpolated functions whereas those in
Figure \ref{fig:delta} simply linearly connect the data points.  The
\dficl\ function was then divided into running bins of $\delta t=0.1$.
For each bin, we shifted the entire \dfficl\ function backward in time
by $\Delta t=0.0-0.1$ in steps of 0.001.  For every value of $\Delta
t$, we calculated the correlation coefficient, $r$, between the
\dficl\ function and the shifted \dfficl\ function, for those points
in the bin of $\delta t$.  The bottom plot of Figure \ref{fig:faint}
shows the maximum value of $r$ calculated in each running bin
($r_{max}$).  The third plot down in Figure \ref{fig:faint} shows the
$\Delta t$ shift associated with each of the $r_{max}$ values ($\Delta
t_{max}$) which is greater than a threshold value of
$r_{thresh}=0.85$.  The value of $r_{thresh}$ was chosen subjectively
to be the value of $r_{max}$ above which the shape of the shifted
\dfficl\ curve reasonably matched that of the \dficl\ curve.

The colored curves in the second plot of Figure \ref{fig:faint}
correspond to the segments of length $\delta t$ of the shifted
\dfficl\ curve, shifted by the $\Delta t_{max}$ of the highest value
of $r_{max}$ within each contiguous set of running bins for which
$r_{max} > r_{thresh}$.  The colored segments of the top plot are the
unshifted segments of \dfficl\ which are shown below.

This analysis illustrates two particularly interesting features.
First, there does seem to exist a real correlation between the
evolution of the ICL at different surface brightnesses, with a time
lag between features seen at high and low surface brightness.
However, while this correlation holds for significant portions of the
clusters' evolution, it does not hold universally for their entire
evolutionary history.  This suggests that while a similar mechanism
may be driving the evolution of both high and low surface brightness
ICL, there is a complex relationship between the two.  Second, the
time lag between high and low surface brightness features increases as
a function of the cluster's evolutionary age.  This is illustrated by
the third plot down in Figure \ref{fig:faint}.  With only one
exception in C2, $\Delta t_{max}$ increases as function of time for
all contiguous sets of $r_{max} > r_{thresh}$.

We suspect that this behavior is due to the changing dynamical scales
on which stripping and ICL production is occurring during the
evolution of the cluster. At early times, the proto-cluster is
basically a collection of smaller groups, which have small physical
dimensions and high density, leading to short dynamical
timescales. When material is stripped in these environments, it
quickly disperses to lower surface brightness, resulting in only a
small time lag between the evolution of high and low surface
brightness material.  As the cluster evolves and grows by accreting
these groups, both the cluster dynamical timescale and the physical
size of the ICL grow; propagating large amounts of luminosity from
high surface brightness to low takes more time and the time lag
between the \dficl\ and \dfficl\ functions grow.  The complexity of
cluster-wide stripping also explains why the correlation between the
ICL and FICL is not universal: it happens only at times where the
stripping and production of ICL is concerted throughout the
volume. For example, the correlation is strong in all simulations at
the beginning, when the galaxies are collapsing in the group
environment throughout the simulation, as well as in clusters C2 and
C3 at later times ($t \approx 0.7$) when there is a cluster-wide
collisional event.

\subsection{ICL Morphology}

As the clusters evolve, not only does the quantity of ICL evolve, but
also the morphology of the ICL features changes dramatically.  As seen
in Figures 5--7, the ICL luminosity at very early times is simply the
outer halos of individual galaxies (see \S \ref{sec:initial} for a
discussion of the simulations at early times).  However, by
$t=0.5-0.6$ each of the clusters show individual sub-groups which have
developed small ICL halos, with very prominent long, thin tidal
features, such as tails, streamers, and filaments.  These narrow
features can be relatively high in surface brightness (\muv$< 29$),
and would be detectable in current broad-band imaging studies --- for
example, see the large scale filaments and plumes visible in Abell
1914 (Feldmeier \etal 2004).  As the evolution of the clusters
progress, the amount of filamentary substructure decreases as the
groups coalesce and the thin tidal streams mix together to form a more
diffuse common envelope at much lower surface brightness, and with
little substructure.  This behavior argues that the morphology of the
ICL, along with its quantity, hold a wealth of information about the
cluster's dynamical history.  We defer a more quantitative
morphological analysis of the ICL to future papers.

\subsection{Radial Distribution of ICL in Evolved Clusters}
We finish the discussion of the ICL structure in our simulated
clusters by focusing on the radial distribution of ICL luminosity at
late times. As shown in Figure \ref{fig:delta}, at late times in their
evolutionary history the fractional ICL luminosity of the clusters is
$\approx 10-15$\%. However, this number is measured over the cluster
as a whole, and can vary significantly as a function of cluster
radius. This variation is shown in Figure \ref{fig:iclfrac}, where the
fractional luminosity is plotted as a function of projected radius. In
each case, the innermost radial bin, at
$R<1.5R_{1/2},$\footnote{$R_{1/2}$ is the half-light radius of the
cluster, $\approx 150-200$ kpc for these clusters.} has a very low ICL
fraction of $\approx 5$\%. This low ICL fraction is due to the fact
that the cluster cores are dominated by luminous galaxies whose high
surface brightness envelopes nearly fill the projected area --- there
is no room for much ICL in the core. Outside the core, the filling
factor of the galaxies drops but the dense environment is still
conducive to tidal stripping, and here we see a sharp increase in the
ICL luminosity fraction ($20-50$\%, depending on the cluster). Further
out from the core, clusters C1 and C3 show a smooth monotonic decrease
in the ICL fraction; in the outskirts of the clusters, galaxies have
not been subjected to the dynamical stripping that contributes so
strongly to the formation of ICL. Cluster C2 shows significantly more
variance, however. This is largely due to how the radial bins sample
the galaxy population, and the presence of bright galaxies in a few of
the radial bins. As a result, the ICL fraction in cluster C2 varies
significantly depending on exactly which radial bin is being
considered --- for example, three bright galaxies all fall in the
third radial bin in cluster C2, suppressing the ICL fraction in that
bin.

The radial profiles shown in Figure \ref{fig:iclfrac} illustrate two
important points. First, on average the ICL fraction shows a radial
gradient, declining in the outer portions of the cluster similar to
the ICL gradients observed in simulations by Murante \etal (2004). In
general, we find that ICL luminosity is centrally concentrated
compared to galactic luminosity, in agreement with the results of both
Murante et al. (2004) and Willman et al. (2004).  Second, the ICL
fraction can vary significantly both between and within clusters,
depending on where and how it is measured. This is due in part to the
relatively low mass of our clusters --- stochastic effects due to
radial binning and the distribution of galaxies will be lessened in
more massive clusters. Nonetheless, this effect demonstrates that
local measurements of the ICL fraction in clusters based on
observations with small areal coverage may not be indicative of the
ICL fraction of the cluster as a whole.

\section{Summary}

In this paper, we have used $N$-body simulations to model the
dynamical evolution of galaxy clusters to study the formation and
evolution of low surface brightness intracluster light. Using an
observational definition for ICL to be at a surface brightness of
\muv$>26.5$ \magsec, we quantify the distribution and luminosity of
the ICL as a function of cluster age, and link the production of ICL
to dynamical events in a cluster's evolutionary history. We also
highlight the importance of the group environment in beginning the ICL
stripping process which is then amplified as groups accrete to form
the cluster.

Our findings suggest that the observed properties of the ICL contain a
wealth of information about the evolutionary history and dynamical
state of the cluster.  In particular, we find that as the cluster
evolves luminous material is transferred from high to low surface
brightness, causing a shift in the cluster's luminosity profile as a
function of surface brightness.  This leads directly to the increasing
fraction of the cluster luminosity found in the ICL.  Because ICL
production is driven by group accretion events within the cluster, the
cluster's ICL luminosity is a sensitive indicator of the cluster's
accretion activity.  We have also shown that very low surface
brightness ICL features are correlated with higher surface brightness
ICL features but lag behind in time, which raises the possibility of
using ICL features of different surface brightness to probe the
clusters' evolution deeper into its dynamical history.  Additionally,
we see a clear evolutionary trend in the morphology of ICL features.
The long, thin tidal features which are so prevalent early in the
cluster's evolution, are destroyed as the cluster accretes, leading to
a more diffuse and amorphous envelope at later times.

In evolved clusters, we find that $\approx 10-15$\% of our clusters'
luminosity is at ICL surface brightness.  These numbers roughly agree
with, but are on the low end of the the $\approx 10-40$\% of cluster
luminosity identified as ICL by other authors using either modeling
(Murante \etal 2004; Willman \etal 2004; Sommer-Larsen \etal 2005) or
direct observations (Ciardullo \etal 2004).  However, it is important
to remember when interpreting these data that they are being compiled
using several different definitions of what constitutes ICL.  Our
definition based on a surface brightness cutoff has the advantages
that it is readily and directly observable, and requires no kinematic
data or modeling of galaxy mass profiles.  However, due to the very
low luminosity of our surface brightness limit of 26.5 \magsec, our
definition of ICL is likely to be rather conservative compared to the
definitions of others.  Our results are in excellent agreement with
the results of Feldmeier et al. (2004) who found isophotal ICL
fractions between $\approx 7-15 \%$ in four Abell clusters when using
a surface brightness limit of \muv=26.5\magsec.  Also, it is worth
noting the result of Murante \etal (2004) which finds that there is a
positive correlation between cluster mass and ICL luminosity fraction.
Our simulated clusters, being of relatively low mass, might then be
expected to have a somewhat smaller fraction of their luminosity as
ICL.

We find that the fraction of cluster luminosity contained in the ICL
tends to increase with time, but does so in a very non-uniform manner.
The production of ICL luminosity appears to be linked to group
accretion events, and is thus a stochastic process intimately tied to
the specific evolutionary history of the cluster. We note that Willman
\etal (2004) claim to see only a weak correlation between accretion
events and the fraction of unbound stars. While some of the
discrepancy may be due to the differing definitions of ICL, another
key difference is in the masses of the clusters studied. Willman \etal
looked at Coma-sized clusters, an order of magnitude more massive than
the clusters studied here. The deep potential well of such massive
clusters will continually and passively strip luminosity from member
galaxies to a greater extent than in these low mass clusters, possibly
leading to markedly different ICL evolution in clusters of
significantly different mass.  However, there exists yet another major
difference between the two analyses, which is the different scales on
which we identify accretion events.  Willman \etal follow only the
large scale accretion of mass into the cluster potential.  Our
analysis, however, is focused on the accretion of and interactions
between small groups of galaxies within the cluster (see a similar
analysis by Gnedin 2003).  We find that the production of ICL is
highly correlated with these small scale events between groups and
sub-groups within the cluster, but not necessarily the mass accretion
history of the cluster as a whole.

In summary, the correlations seen here between the properties of the
ICL and the evolution of galaxy clusters means that measurements of
the ICL provide a powerful new tool for studying the dynamical history
of clusters.  Continuing investigations into low surface brightness
ICL features --- including studies of their luminosity, morphology,
and kinematics, as well as the physical linkage to individual galaxies
--- should reveal unprecedented information about the evolution of
galaxy clusters and groups, and the processes affecting the evolution
of the galaxies within them.

\acknowledgments

CSR appreciates support from the Jason J. Nassau Graduate Fellowship
Fund.  JCM acknowledges research support from the NSF through grant
ASTR 98-76143 and from a Research Corporation Cottrell Scholarship.

\newpage
\input{tab1.tex}

\begin{figure}
\caption{A simulated image of Cluster C2 at $t=0.67$, color coded by
  $V$-band surface brightness, \muv.  Black represents all luminosity
  at \muv$<26.0$ \magsec.  Each other color represents a bin of one
  \magsec: red is \muv$=26.0-27.0$, orange is \muv$=27.0-28.0$, etc.
  This same color scheme will be used in all simulated images shown in
  this paper.  The distance scale in the upper right indicates the
  length of one megaparsec in physical units.
\label{fig:example}}
\end{figure}

\begin{figure}
\plotone{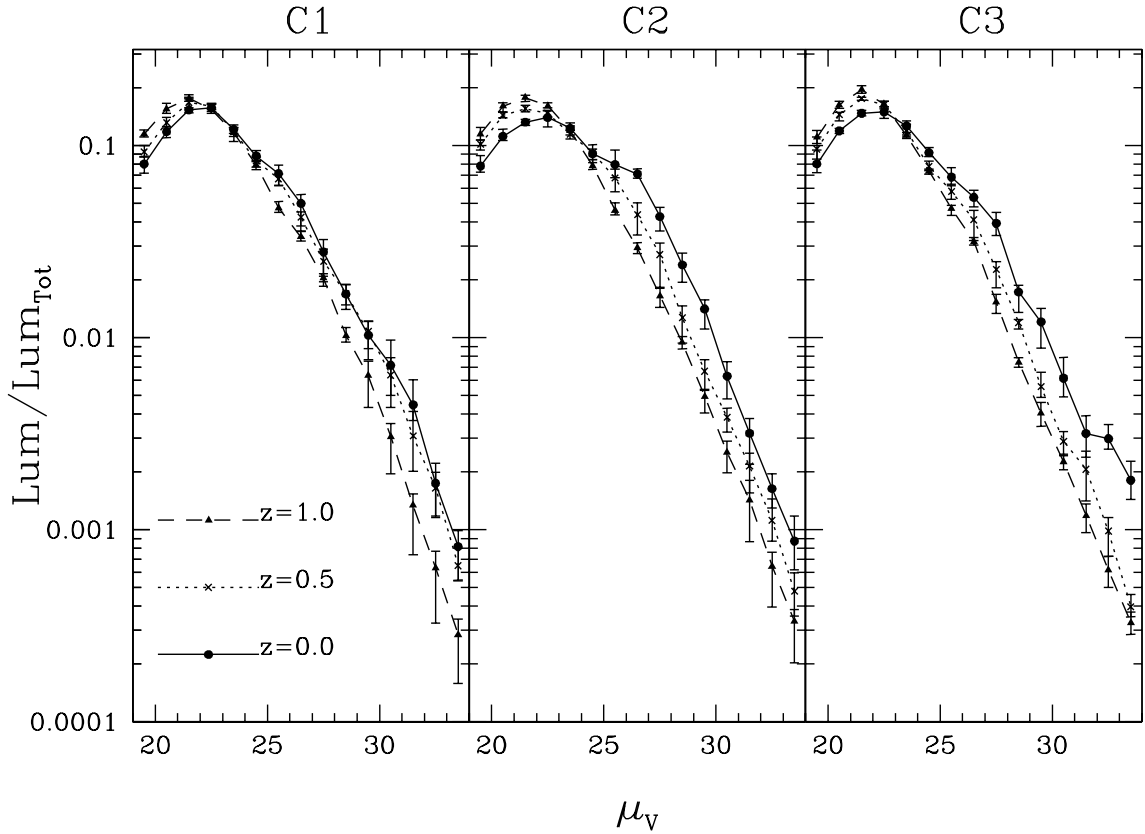}
\caption{The fraction of cluster luminosity contained in 1 \magsec\
  bins of \muv\ for each of the three clusters at three different
  timepoints in their evolution.
\label{fig:rotlum}}
\end{figure}

\begin{figure*}
\centerline{\includegraphics[scale=0.99,angle=0]{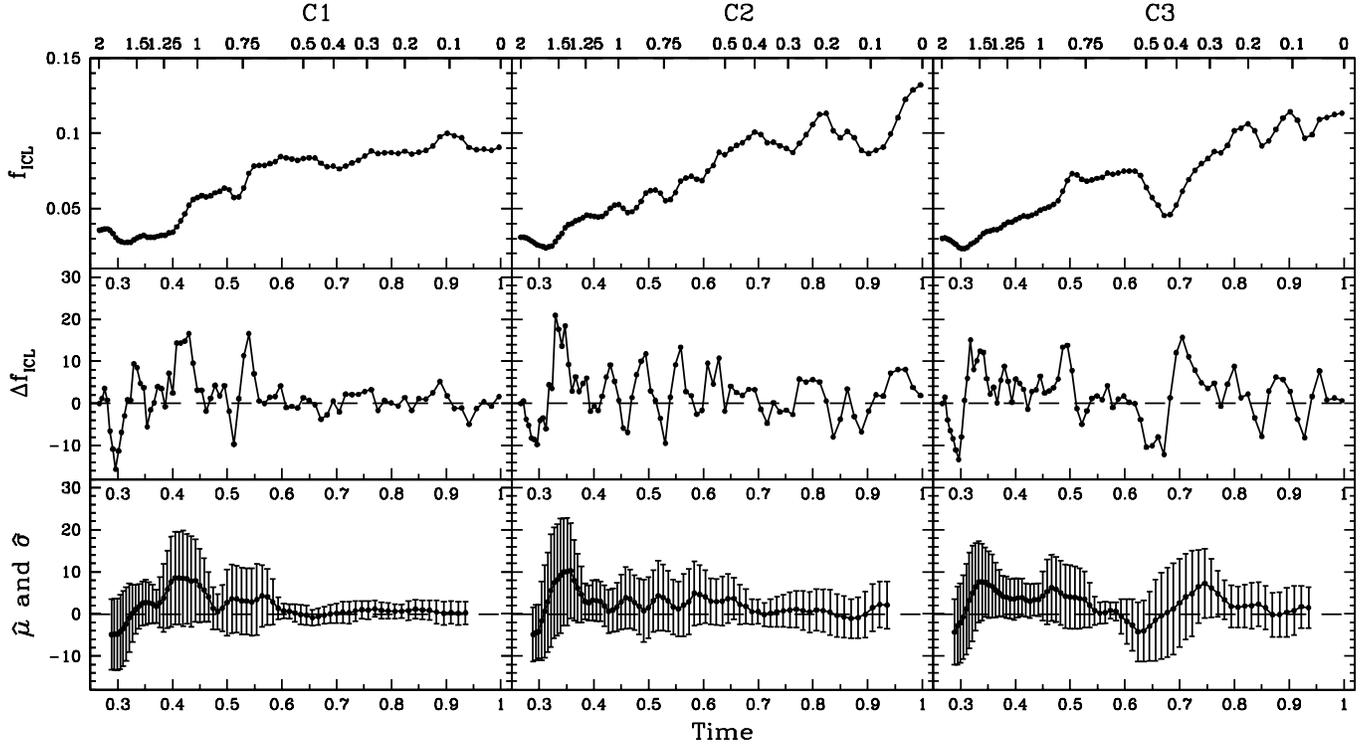}}
\caption{\emph{Top}: The fraction of the luminosity in each cluster
  which is at ICL surface brightness (\muv$>26.5$ \magsec) as a
  function of time (\ficl).  The top axis shows the corresponding
  nominal redshift.  \emph{Middle}: The fractional change in ICL
  luminosity per unit time as a function of time (\dficl).
  \emph{Bottom}: The running mean ($\hat{\mu}$) with standard
  deviation ($\hat{\sigma}$) error bars, using a running bin of ten
  time points, for the \dficl\ function.  The time is the average of
  the timepoints in each bin.
\label{fig:delta}}
\end{figure*}

\begin{figure*}
\caption{Simulated images of the evolution of Cluster C1 color coded
  by $V$-band surface brightness, \muv.  Black represents all
  luminosity at \muv$<26.0$ \magsec.  Each other color represents a
  bin of one \magsec: red is \muv$=26.0-27.0$, orange is
  \muv$=27.0-28.0$, etc.  The color scale is identical to that used in
  Figure \ref{fig:example}.  The distance scale in the upper left
  indicates the length of one megaparsec in physical units.
\label{fig:c1}}
\end{figure*}

\begin{figure*}
\caption{Simulated images of the evolution of Cluster C2 color coded
  by $V$-band surface brightness, \muv.  Black represents all
  luminosity at \muv$<26.0$ \magsec.  Each other color represents a
  bin of one \magsec: red is \muv$=26.0-27.0$, orange is
  \muv$=27.0-28.0$, etc.  The color scale is identical to that used in
  Figure \ref{fig:example}.  The distance scale in the upper left
  indicates the length of one megaparsec in physical units.
\label{fig:c2}}
\end{figure*}

\begin{figure*}
\caption{Simulated images of the evolution of Cluster C3 color coded
  by $V$-band surface brightness, \muv.  Black represents all
  luminosity at \muv$<26.0$ \magsec.  Each other color represents a
  bin of one \magsec: red is \muv$=26.0-27.0$, orange is
  \muv$=27.0-28.0$, etc.  The color scale is identical to that used in
  Figure \ref{fig:example}.  The distance scale in the upper left
  indicates the length of one megaparsec in physical units.
\label{fig:c3}}
\end{figure*}

\begin{figure*}
\caption{Simulated images of a short section of the evolution C2 at
  high time resolution color coded by $V$-band surface brightness,
  \muv.  Black represents all luminosity at \muv$<26.0$ \magsec.  Each
  other color represents a bin of one \magsec: red is
  \muv$=26.0-27.0$, orange is \muv$=27.0-28.0$, etc.  The color scale
  is identical to that used in Figure \ref{fig:example}.  The distance
  scale in the upper left indicates the length of one megaparsec in
  physical units.  The arrow points to the small galaxy group which
  moves through the cluster core (see Section \ref{sec:hires} for
  details).
\label{fig:hires}}
\end{figure*}

\begin{figure}
\plotone{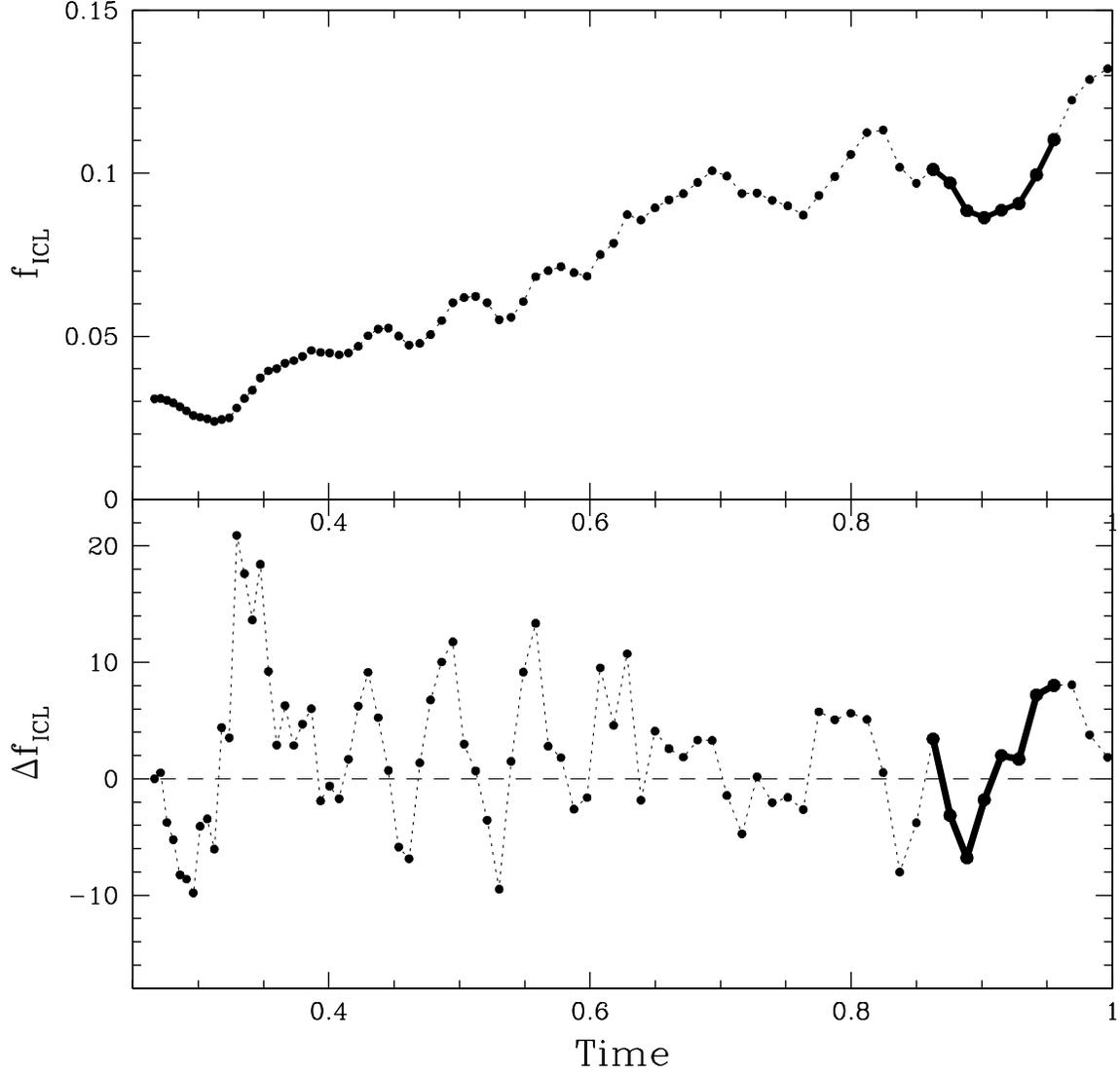}
\caption{The \ficl\ and \dficl\ plots for C2.  The bold section
  indicates the section used in the high time resolution analysis of
  \S \ref{sec:hires}.
\label{fig:hires_plot}}
\end{figure}

\begin{figure*}
\caption{\emph{Top}: The \dficl\ function for each cluster (closed
  circles) and the \dfficl\ function (open squares) which is simply
  the \dficl\ function for luminosity at \muv$>30.0$.  The curves
  follow a third order polynomial fit to the data.  The colored curves
  are those sections of the \dfficl\ which are highly correlated with
  sections of the \dficl.  \emph{Middle Top}: The \dficl\ function,
  with the highly correlated segments of the \dfficl\ function shifted
  in time by $\Delta t_{max}$.  \emph{Middle Bottom}: $\Delta
  t_{max}$, or the time shift which results in the highest
  correlation, for the highly correlated segments of the \dfficl\
  function.  \emph{Bottom}: The highest value of the correlation
  coefficient, $r$, for each bin of the \dfficl\ function resulting
  from shifting \dfficl\ backward in time by $\Delta t < 0.1$.
\label{fig:faint}}
\end{figure*}

\begin{figure}
\plotone{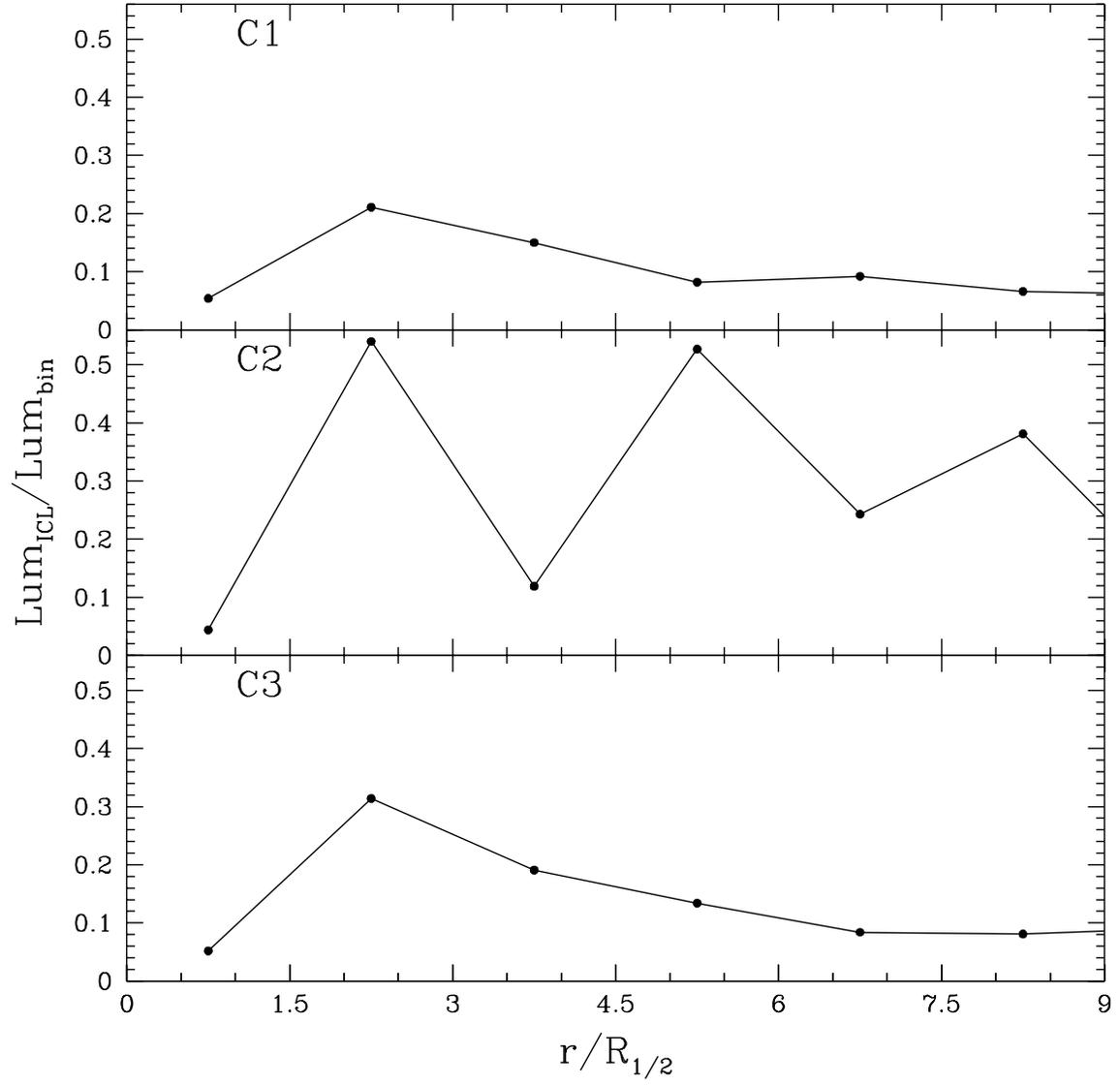}
\caption{The fraction of ICL luminosity as a function of projected
radius in each of the three simulated clusters. \label{fig:iclfrac}}
\end{figure}

\end{document}

%% file: tab1.tex
\begin{deluxetable}{lrrrr}
  \tablecolumns{5}
  \tablewidth{0pc}
  \tablecaption{Properties of the Simulated Clusters \label{tab:ctable}}
  \tablehead{
    \colhead{Cluster} & \colhead{$R_{200}$} & \colhead{$M_{200}$} &
  $N_{gal}$ & $N_{gal}$ \\
    \colhead{} & \colhead{(Mpc)} & \colhead{$(M_{\odot})$} &
    \colhead{($z=2$)} & \colhead{($z=0$)}
  }
  \startdata
    C1 & 0.95 & $9.3\times10^{13}$ & 121 & 75  \\
    C2 & 0.94 & $9.1\times10^{13}$ & 143 & 89  \\
    C3 & 0.98 & $1.0\times10^{14}$ & 164 & 106 \\
  \enddata
\end{deluxetable}